# Magnetoelectric coupling in polycrystalline FeVO$_4$

Bohdan Kundys[†], Christine Martin[*] and Charles Simon[*]

[†]*Commissariat of Atomic Energy, Solid State Phys. Division (SPEC), CEA Saclay, 91191 Gif-Sur-Yvette Cedex, France*
[*]*Laboratoire CRISMAT, CNRS UMR 6508, ENSICAEN, 6 Boulevard du Maréchal Juin, 14050 Caen Cedex, France*

We report coupling between magnetic and electric orders for antiferromagnetic polycrystalline FeVO$_4$ in which magnetism-induced polarization has been recently found in noncollinear antiferromagnetic state below the second antiferromagnetic phase transition at $T_{N2} \approx 15.7$K. In this low symmetry phase space group $P\bar{1}$, the magnetic field dependence of electric polarization evidences a clear magnetoelectric coupling in the noncollinear spin-configured antiferromagnetic phase. The discontinuity of magnetodielectric effect observed at the vicinity of the polar to nonpolar transition evidences competition between different magnetodielectric couplings in the two different antiferromagnetic states. The existence of thermal expansion anomaly near $T_{N2}$ and magnetostriction effect support magnetoelastically mediated scenario of the observed magnetoelectric effect.

Renewed attention to the magnetoelectric effects in solids[1–3] has stimulated the exploration of magnetoelectric materials as well as reinvestigation of previously known compounds. Because underlining mechanism of magnetism-induced electric polarization in solids remains still debated magnetoelectric materials in which only one ion is magnetic are particularly important for fundamental understanding in depth magnetoelectric coupling. In this respect the recent report on magnetic and electric orders coexistence in FeVO$_4$ (space group $P\bar{1}$) has added data for magnetic multiferroic materials design.[4] Here, we report that FeVO$_4$ not only possesses magnetic and electric orders at the same temperature region below the second antiferromagnetic phase transition ($T_{N2} \approx 15.7$K) but also reveals a clear coupling between magnetic and electric orders. Fact that the electrical polarization in majority of recently discovered materials[5-8] lies in the plane of the spin spiral, but in a direction that is perpendicular to its propagation vector has stimulated studies in noncollinear antiferromagnetic crystals in order to demonstrate magnetic field induced switching of electric polarization. While this approach is successful in describing magnetism-induced electric polarization in many samples, it appears recently that electric polarization may exist irrespectively of spin-chirality propagation direction in DyMnO$_3$ (Ref. 9) and in CuFe$_{1-x}$Al$_x$O$_2$.[10] These facts suggest that for some compounds the spin-chirality anisotropy-dependent rule (spin current model)[11] is not critical limitation for magnetic field induced polarization, but rather spontaneous magnetostriction-caused lattice deformation at the spin reorientation is responsible for nonpolar to polar transition. Indeed, spin reorientations are known to be accompanied with structural deformations induced by either magnetic field[12-16] or temperature[7,17] in many compounds. Taking into account this fact and the complexity that is linked to crystal growth and cutting along the desired crystallographic directions, it becomes also very interesting to study polycrystalline samples instead of single crystals ones that are more time consuming in preparation. Additionally fact that phenomenon of magnetic field induced switching of the electric polarization from one crystallographic direction to another has been found in rare-earth manganites[18,19] and in MnWO$_4$ (Refs. 8, 20, and 21) suggests that electric polarization may exist along specific crystallographic direction only at the given temperature in a zero magnetic field and strong magnetic field is needed to switch polarization between the axis. For this reason polycrystalline samples also offer advantage in more rapid magnetoelectric sample characterization for speeding up the search for magnetic multiferroics. Indeed, magnetism-induced polarization has been recently reported for polycrystals of CuCrO$_2$,[22] YBaCuFeO$_5$,[23] and CuFe$_{0.95}$Rh$_{0.05}$O$_2$.[24]

Polycrystalline sample of FeVO$_4$ with dimensions of 0.58x3.6x2.2 mm$^3$ was used in this study. The magnetization measurements were carried out using physical property measurement system (PPMS Quantum Design) system. The dielectric measurements were done in PPMS cryostat using an Agilent 4248A RLC bridge at 100 mV oscillation voltage amplitude. Polarization was measured using Keithley 6517A electrometer possessing automatic current integration facility for more precise charge measurements. The magnetic field was applied perpendicular to the direction of the electric field. Silver paste was used to make electrical contacts to the sample. The magnetostriction and thermal expansion were measured using modified capacitance dilatometer technique[25] in PPMS cryostat. To keep the same configuration of experiment as for magnetopolarization measurements magnetic field was applied perpendicular to the direction in which magnetostriction was measured. The magnetic properties of single crystalline phase of FeVO$_4$ can be found elsewhere.[4] The polycrystaline FeVO$_4$ as well, undergoes successive magnetic transitions at 23.3K ($T_{N1}$) and 15.7K ($T_{N2}$) (Fig. 1(a)). While on lowering temperature collinear[4] antiferromagnetic transition at $T_{N1}$ shows no anomaly in dielectric properties a clear peak in dielectric permittivity is accompanied by the second noncollinear transition at $T_{N2}$ (Fig. 1(b)). The temperature position of the anomaly in dielectric permittivity at $T_{N2}$ was found to be independent of measurement frequency (Fig. 1(b)) implying its strong correlation with magnetic order. Furthermore, the both dielectric peak correlation with dielectric loss and low loss level itself (Fig. 1(b) (inset)) are common features for polar transitions.[26] To check if electric polarization appears in this temperature region we have measured electric polarization with an electrometer.

[†]Electronic mail: kundys@.gmail.com (Bohdan Kundys).

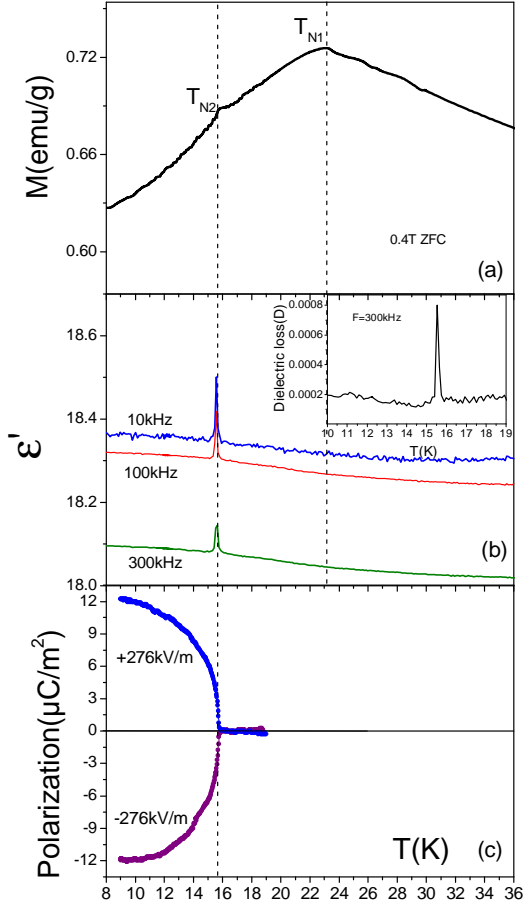

FIG. 1. (Color online) (a) ZFC (0.4 T) magnetization, (b) dielectric permittivity at different frequencies, and (c) electric polarization of FeVO$_4$ as a function of temperature. Inset shows dielectric loss as a function of frequency at 300 kHz.

Since the magnitude of spontaneous electric polarization can be small in many magnetic and electrically polar compounds we have previously cooled sample from 35 K down to 9 K in electric field of 276 kV/m. Next the electric field was removed and time dependence of the forced polarization has been measured during 3000s to ensure polarization stability. Figure 1(c) confirms existence of electric polarization for $T \approx 15.7$ K (i.e.,$T_{N2}$). A sign of the polarization may be thermodynamically switched to the positive or negative values when corresponding (negative or positive) electric field cooling procedure is applied. The peak in dielectric anomaly reveals strong magnetic field dependence (Fig. 2) and its position shifts toward lower temperatures with external magnetic field application (Fig. 2 (inset)). This behavior is very similar to that observed in other polar and magnetic samples[19,27] and therefore may analogously imply magnetoelectric coupling. It has to be also noted that the difference between zero magnetic field dielectric permittivity curve and dielectric permittivity curve taken at 4T, that is proportional to electromagnetic susceptibility of the sample versus temperature, reveals discontinuity around 15.6 K (i.e.,$T_{N2}$ )(Fig. 3). This behavior may be a result of competition between two antiferro magnetic phases in magnetodielectric coupling at the vicinity of polar-nonpolar transition as was previously observed in other magnetoelectric samples.[24,28] Fact that, magnetodielectric effect increases as temperature approaches the transition to nonpolar state as shown in Fig.3(see electrically polar region $T$ 15.6 K), was previously observed for polar antiferromagnets Cr$_2$O$_3$ (Ref. 29) and YBaCuFeO$_5$.[23] This seems to be a common feature for many compounds if one takes into account a simple phenomenological model.[30] However, the observed discontinuity at the transition is a result of magnetic phase competition and confirms that magnetodielectric effects may indeed be present even in paraelectric antiferromagnetic region at the vicinity of the polar to nonpolar transition in compounds with competing magnetic orderings. In order to observe the magnetoelectric coupling we have previously cooled sample from 35 K down to 9K in electric field of 276 kV/m. Next electric field was removed and time dependence of the forced polarization has been measured during 3000s again to ensure polarization stability. We have then heated sample thereafter from 9 to 10 K followed by time-dependent measurements of the forced electric polarization with simultaneous magnetic field oscillation of 14T amplitude. This procedure was chosen to examine polarization stability over time. As magnetic field reaches its maximum value of 14 T, the polarization reaches its minimum (Fig.4). The oscillation of the magnetic field of ±14 T induces the oscillation of electric polarization with the maximum change of about 3.7 C/m$^2$ at the double frequency, evidencing the existence of a clear magnetoelectric coupling in the sample. More interestingly, there is a correlation between magnetopolarization (Fig. 4) and magnetoelastic (Fig. 5) effects in the sample. Namely, maximum of magnetic field induces a maximum of lattice deformation that corresponds to minimum of polarization. This effect must be related to the suppression of electrically favorable spin configuration by the external magnetic field in agreement with decrease in peak in the dielectric permittivity (corresponding to polar-nonpolar transition) under external magnetic field application (Fig. 2). These results suggest that magnetoelasticity may play dominant role of the observed magnetoelectric effect. If this is true, the electric polarization should appear as a result of thermodynamically driven spontaneous magnetostriction effect[17] at the spin-reorientation transition. Indeed anomaly in thermal expansion in our sample has been found at $T_{N2}$ (Fig. 6) in zero magnetic field. Fact that it disappears with magnetic field application additionally confirms the importance of magnetoelastic effects in the observed magnetoelectric coupling. It is also feasible that some reduced thermal expansion anomaly may still exist under the applied magnetic field, this, however, may be out of sensitivity limit of the instrument. It also has to be noted that cooling under magnetic field from collinear to noncollinear antiferromagnetic state may naturally freeze the collinear antiferromagnetic arrangement. Consequently, this prevents otherwise thermodynamically favorable noncollinear and polar state to be formed. Because exchange interactions depend on distances between magnetic interaction centers (that are now magnetically frozen collinearly) thermal expansion tends to disappear under magnetic field more rapidly comparing to zero magnetic field cooling procedure. This is probably why in P(H) measurements (done in zero magnetic field cooling (Fig. 4)) magnetic field is not so strong to destroy electric polarization completely because spins were not collinearly frozen.

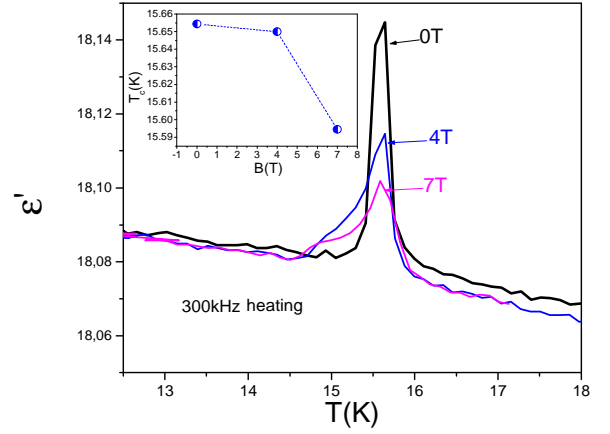

FIG. 2. (Color online)Dielectric permittivity of FeVO$_4$ at 300kHz as a function of temperature at different magnetic fields recorded on heating. Inset shows evolution of $T_c$ as a function of magnetic field.

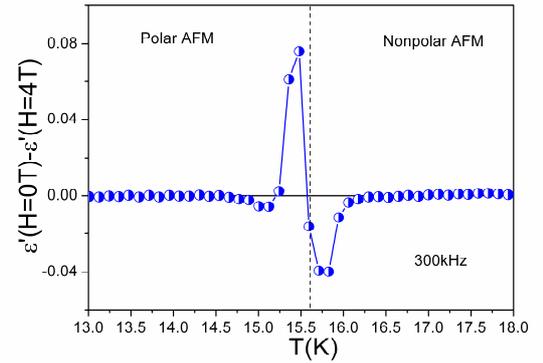

FIG. 3. (Color online) Difference between zero magnetic field dielectric permittivity curve and dielectric permittivity curve taken at 4 T (see Fig. 2) as a function of temperature.

Our efforts to demonstrate 180° electric dipole reversibility of the electric polarization with electric field of 430 kV/m (i.e., to demonstrate ferroelectricity) at different magnetic fields and temperatures in the sample were unsuccessful, despite the fact that cooling down with a different direction of the electric field indeed switches the dipoles thermodynamically by certain angle (Fig. 1(c). This however cannot be taken as prove of ferroelectricity. Taking into account that the symmetry group of our sample is the $P\bar{1}$ , it is rather difficult to switch polarization by 180° with electric field as it is predicted by Landau analysis. Thus, it can be assumed that our sample is rather just electrically polar compound than ferroelectric coercive force is too big.

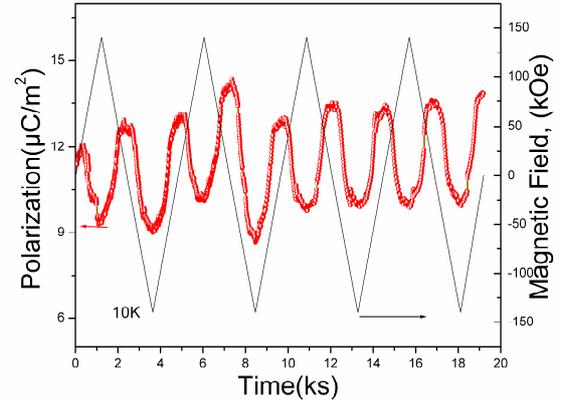

FIG. 4. (Color online)Time dependence of the forced electric polarization of FeVO$_4$ (left scale) with simultaneous magnetic field change (right scale) at 10 K.

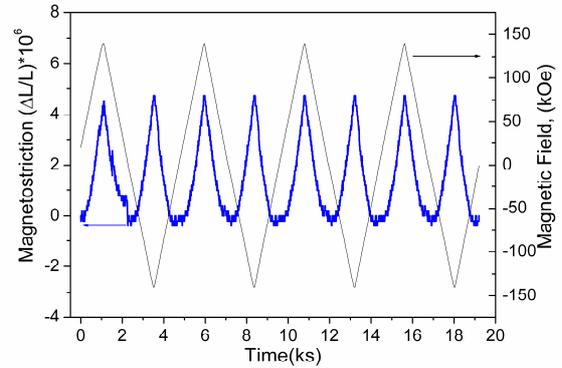

FIG. 5. (Color online)Time dependence of the elastic deformation of FeVO$_4$ (left scale) with simultaneous magnetic field change (right scale) at 10 K.

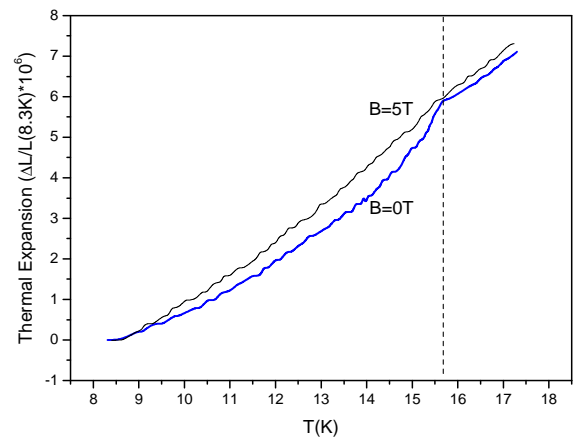

FIG. 6. (Color online)Thermal expansion of FeVO$_4$ measured on cooling at 0 and 5 T magnetic fields.

In summary, it was found that FeVO$_4$ is a magnetoelectric compound in which both electric polarization and antiferromagnetism coexist in a same temperature region and couple. The appearance of incommensurate antiferromagnetic phase determined from the temperature dependence of magnetization and neutron scattering data is accompanied with classical anomaly in dielectric permittivity. The clear correlation between magnetostriction and magnetopolarization effect implies dominating contribution of magnetoelasticity in the observed magnetoelectric coupling. The magnetic field dependent thermal expansion anomaly at $T_{N2}$ additionally supports this assumption. Finally, our results show that even in polycrystals, the averaging of the electric polarization by using polycystaline samples is not a redhibitory limitation, but can even provide an advantage over the crystals in more rapid sample characterization. Such a study should help in the screening of other similar magnetic materials in the search of magnetic multiferroics and in understanding of microscopic origin of magnetoelectric coupling.